\documentclass[fleqn,usenatbib]{mnras}

\usepackage{newtxtext,newtxmath}
\usepackage[T1]{fontenc}
\usepackage{ae,aecompl}

\usepackage{enumitem}
\usepackage{pgfplots}
\pgfplotsset{compat=1.18}
\usepackage{comment} 
\usepackage{hang}
\usepackage{array}

\usepackage{graphicx}	% Including figure files
\usepackage{amsmath}	% Advanced maths commands
\title{Observations concerning the extinction of humanity and exocivilizations}

\author[Darren J. Dougan]{
Darren J. Dougan,$^{1}$\thanks{E-mail: darren.dougan@bqi.com.au}
\\
$^{1}$Big Questions Institute, 55 Holt St Surry Hills, Sydney NSW 2010 Australia\\
}

\date{Accepted XXX. Received YYY; in original form ZZZ}

\pubyear{2025}

\begin{document}
\label{firstpage}
\pagerange{\pageref{firstpage}--\pageref{lastpage}}
\maketitle

\begin{abstract}
The Great Filter hypothesis is an extension of the Fermi Paradox: \textit{"If life is so common in the universe, why don't we see it?"}  The Great Filter theory posits there are multiple obstacles or "Filters" that life must pass through which ultimately sifts out intelligent life.\\

\noindent This paper identifies a new filter - \textbf{depopulation}.  As a species advances and reaches the top of the food chain on its planet, Darwinian evolution selects the species to breed fewer offspring due to a lack of predation.  As the species evolves intelligence, this leads to medicines and most notably contraception, enabling the species to reduce infant mortality while controlling reproduction.  Finally, economic, social and educational factors add to the conscious decision of intelligent life to slow reproduction.  These factors are currently contributing to a human global population peak later this century with subsequent population collapse in less than 500 years.  Noting that population growth (\textbf{\textit{and decline}}) is exponential, our modelling forecasts human extinction thresholds being tested some time after the year 2500.  There is no reason to assume depopulation dynamics (exodepopulation) would not apply to exocivilizations (exodemography), thus providing a possible resolution to the Fermi Paradox.\\ 

\noindent Furthermore, as machines and AIs inevitably supplement humans as depopulation accelerates, the Fermi paradox can be restated as \textit{"Why don't we see machines and AIs colonizing the galaxy?"}  A plausible answer is machines \textbf{will not become conscious} and will continue to act only as tools; tools that will cease operating once humanity is extinct.  The Fermi Paradox can then be restated as \textit{"Machines will not become conscious, otherwise we would see them colonizing the universe"}.
\end{abstract}


%%%%%%%%%%%%%%%%%%%%%%%%%%%%%%%%%%%%%%%%%%%%%%%%%%

%%%%%%%%%%%%%%%%% BODY OF PAPER %%%%%%%%%%%%%%%%%%

\section{Introduction}
The Great Filter hypothesis \cite{Hanson1998} suggests there are large barriers and obstacles to intelligent life evolving and colonising the universe.  The type of filters may include such factors as: 

\begin{enumerate} [label=\arabic*., leftmargin=2em]
    \item The right star system (including organics \& habitable planets)
    \item Reproductive molecules (e.g. RNA)
    \item Simple (prokaryotic) single-cell life
    \item Complex (archaeatic and eukaryotic) single-cell life
    \item Sexual reproduction
    \item Multi-cell life
    \item Tool-using animals with intelligence
    \item A civilization advancing toward the potential for a colonization explosion (where we are now)
    \item Colonization explosion
\end{enumerate}

The Great Filter is the filter that extinguishes life.  When applied to humanity, it may be that the Great Filter is in front of us in time, which means we are likely to become extinct sometime in the future.  However, if the Great Filter is behind us, then we may survive extinction.  

The Great Filter hypothesis is analogous to the Drake-Sagan formula \cite{Drake1975} which attempts to calculate the number of exocivilizations in the galaxy at any given time.  In both cases, the absence of observable life in the universe, and most notably our galaxy, indicates that life is extremely rare, and the filter may happen early in life's evolution.

Notwithstanding, humanity appears to have passed through significant filters (at a minimum all seven listed above).  \cite{Carter1983Anthropic} used statistical methodologies to estimate that life must have passed through at least two "hard steps"; these steps being extremely improbable.  Carter's anthropic argument suggested that humanity is 80\% through its lifetime and extinction lies in the future.  

Like Carter, this paper hypothesizes that the Great Filter is ahead of humanity and that filter is \textbf{\textit{depopulation}}.

\section{Depopulation Models}

\subsection{Introduction}

Population forecasting models are well understood and have been applied across numerous disciplines including biology, physics and the social sciences. A brief introduction is given here.

Let the number of individuals in a population be $n$.  $n$ varies with time $t$ so $n(t)$ represents the population function.  $\frac{dn}{dt}$ represents the growth rate of the population and $\frac{1}{n}\frac{dn}{dt}$ represents the per capita growth rate.  If the per capita birth rate is given by $b$ and the per capita death rate is given by $d$, then, 

\begin{eqnarray}
    \frac{1}{n}\frac{dn}{dt} & = & b - d \\
    & = & g
\end{eqnarray}
where $g$ is the compound annual growth rate (CAGR) of the population. This can also be written as, 

\begin{eqnarray}
    \frac{dn}{dt} & = & g n 
\end{eqnarray}

\noindent Equation (3) is a differential equation which has the solution,

\begin{eqnarray}
    n(t) & = & n_0e^{gt}
\end{eqnarray}

\noindent where $n_{0}$ is the number of individuals at the start of the process.  

This model is exponential and has limitless growth.  It also assumes constant growth rates.  To make more realistic, one adds constraints including the carrying capacity of the environment, denoted $K$. 

\begin{eqnarray}
    \frac{dn}{dt} & = & gn \bigl(1 - \frac{n}{K} \bigr) 
    % & = & rN - \frac{rN^2}{K}
\end{eqnarray}
This equation is called the \textit{Logistic Equation} and is now a quadratic equation of the population growth rate.  Clearly, when the population size equals the carrying capacity $n = K$, $\frac{dn}{dt} = 0$ and growth in the systems stops.  This assumes a fixed carrying capacity $K$.  

\subsection{Model Application to Humanity}

Population models have been applied to human circumstances as far back as Benjamin Franklin's observations on the growth of the American population \cite{Franklin1755Observations}.  Franklin calculated a doubling of the pre-revolution American population every 20 years assuming a birth rate of 8 children per woman and an infant mortality rate of 4 children per woman.

Since then, most population growth analysis has been focused on growing human populations with inherent resource constraints.  \cite{BasenerRoss2005} developed a model for Easter Island which tracked both the population $n$ and a resource $r$ directly related to the island’s human carrying capacity (see also \cite{Frank2018}, \cite{Safuan2012_predator_prey}).  The resource was renewable and had its own carrying capacity, but was also consumed by humans.  The system was modeled as below:
\begin{eqnarray}
    \frac{dn}{dt} & = & gn \bigl(1 - \frac{n}{r} \bigr) 
\end{eqnarray}

\begin{eqnarray}
    \frac{dr}{dt} & = & Cn \bigl(1 - \frac{n}{k} \bigr) - Hn
\end{eqnarray}
where $C$ is the growth rate of the resource, $k$ is its carrying capacity and $H$ is the rate of human consumption.  The modeling showed a relatively comparable collapse in the Easter Island population as the carrying capacity of the island was exceeded and the resource exhausted.  However, it should be noted that in advanced economies, the carrying capacity of the environment is typically not exceeded ($n < K$).  Japan for instance, has declining population with excess resources.  The population of Japan is not short of air, water, food, shelter or economic means, unlike the Easter Island inhabitants yet population is in decline.  

Turning to humanity as a whole, what if $d > b$?  Then $g$ is negative and inverse exponential growth occurs, 

\begin{eqnarray}
    n(t) & = & n_0e^{-gt}
\end{eqnarray}

Human replacement population rates assume 2.1 births per female to maintain a current population (the additional 0.1 reflects infants dying before reaching child bearing age).  This is known as the Total Fertility Rate (TFR).  TFR = 2.1 equates to $g=0$.  If TFR falls below 2.1, then $g$ becomes negative, if TFR is greater than 2.1, then $g$ is positive.  

\begin{figure}
    \centering
    
    \begin{tikzpicture}
        \begin{axis}[
            width=8cm,
            height=8cm,
            xlabel={Year},
            xticklabel style={
                /pgf/number format/.cd,
                fixed,
                precision=0,
                1000 sep={}
            },
            ylabel={Fertility Rate (Births per Woman)},
            yticklabel style={
                /pgf/number format/.cd,
                fixed,
                fixed zerofill, % ensures trailing zeros are shown
                precision=1      % one decimal place
            },
            grid=major,
            xmin=1960, xmax=2023,
            ymin=1.5, ymax=5.5,
            xtick={1960,1970,1980,1990,2000,2010,2020},
            ytick={2.0,2.5,3.0,3.5,4.0,4.5,5.0,5.5},
            thick,
        ]

        \addplot[
            color=blue,
            mark=none
        ]
        coordinates {
        (1960,4.7) (1961,4.6) (1962,5.0) (1963,5.3) (1964,5.1) (1965,5.1)
        (1966,5.0) (1967,4.9) (1968,5.0) (1969,4.9) (1970,4.8) (1971,4.7)
        (1972,4.5) (1973,4.4) (1974,4.3) (1975,4.1) (1976,4.0) (1977,3.8)
        (1978,3.8) (1979,3.8) (1980,3.7) (1981,3.7) (1982,3.7) (1983,3.6)
        (1984,3.6) (1985,3.5) (1986,3.5) (1987,3.5) (1988,3.4) (1989,3.3)
        (1990,3.3) (1991,3.1) (1992,3.0) (1993,3.0) (1994,2.9) (1995,2.9)
        (1996,2.8) (1997,2.8) (1998,2.7) (1999,2.7) (2000,2.7) (2001,2.7)
        (2002,2.7) (2003,2.6) (2004,2.6) (2005,2.6) (2006,2.6) (2007,2.6)
        (2008,2.6) (2009,2.6) (2010,2.6) (2011,2.6) (2012,2.6) (2013,2.6)
        (2014,2.5) (2015,2.6) (2016,2.5) (2017,2.5) (2018,2.5) (2019,2.5)
        (2020,2.4) (2021,2.3) (2022,2.2) (2023,2.2)
        };

    \end{axis}
    \end{tikzpicture}

    \caption{World Fertility Rate (1960--2023)}
    \label{fig:TFRHistory}
\end{figure}
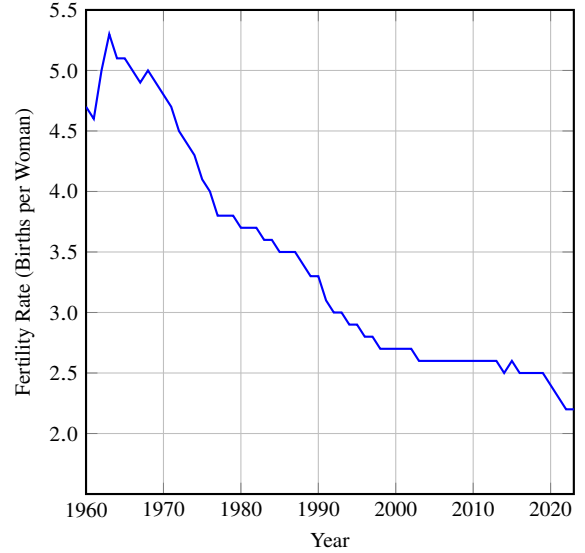

In 1963, global TFR peaked at 5.3 but by 2023, the fertility rate had fallen to 2.2. \cite{WorldFertility2023}.  This is shown in Figure~\ref{fig:TFRHistory}.  Since 2016, TFR has fallen 15\% each year, or a rate reduction of approximately 0.5 each year.  Extrapolating to 2026, it seems likely that global TFR is lower than replacement value 2.1.

Tables \ref{tab:TFRIncome}, \ref{tab:TFRArea}, \ref{tab:TFRCountry} (\cite{WorldFertility2023}) show fertility rates by income, area and country.  On a granular level, we see countries now experiencing negative population growth.  China's birth rate in 1963 was 7.5, but has dramatically fallen to 1.0 by 2023 due to increasing wealth and the one child policy.  China's population declined by 3.4 million people in 2025 while it is forecast that China's population will halve by 2100.  India's TFR is now 1.98, below replacement value, after peaking at 6.0 in 1963.  So the two most populous countries in the world have birth rates below replacement.

\begin{table}
	\centering
	\caption{TFR by income}
	\label{tab:TFRIncome}
	\begin{tabular}{lc} % four columns, alignment for each
		\hline
		Income & TFR \\
		\hline
		High & 1.4 \\
		Upper middle  & 1.5 \\
		Middle & 2.1 \\
		Low and middle & 2.3 \\
        Lower middle & 2.6 \\
        Low & 4.7 \\
        \hline
	\end{tabular}
\end{table}

\begin{table}
	\centering
	\caption{TFR by geographic area}
	\label{tab:TFRArea}
	\begin{tabular}{lc} % four columns, alignment for each
		\hline
		Area & TFR \\
		\hline
		European Union & 1.4 \\
		East Asia & 1.6 \\
        Europe and Central Asia   & 1.6 \\
		North America   & 1.6 \\
		Latin America  & 1.8 \\
		South Asia & 2.0 \\
        Arab world & 3.1 \\
        Sub Saharan Africa & 4.4 \\
        \hline
	\end{tabular}
\end{table}

\begin{table}
	\centering
	\caption{TFR by selected country}
	\label{tab:TFRCountry}
	\begin{tabular}{lcc} % four columns, alignment for each
		\hline
		Country & Growth rate (\%) & TFR\\
		\hline
		Japan & -0.5\% & 1.2\\
		Greece & -0.4\% & 1.3\\
        China & -0.1\% & 1.0\\
        Australia & 1.7\% & 1.5\\
		USA & 1.0\% & 1.6\\
		Brazil & 0.5\% & 1.6\\
		India & 0.9\% & 2.0\\
		Indonesia & 0.7\% & 2.1\\
		Pakistan & 1.9\% & 3.6\\
		Nigeria & 2.5\% & 4.5\\
		\textbf{WORLD} & \textbf{1.0\%}& \textbf{2.2}\\\hline
	\end{tabular}
\end{table}

The reason for this steady reduction in fertility rates is not consistent across geographies, cultures or incomes.  The prevailing view of correlation to economic wealth is not supported in areas like Calcutta with a TFR of 0.6, or South America which is under 2.0.

Affordability concerns, lower rates of marriage, digitization and less socialization, future environmental and geopolitical concerns, and higher levels of women’s education and career participation are all possible explanations.  Access to contraception is also a key factor.

Statistics show women are having children later in life which reduces the number of births.  One possible consistent explanation across all examples is the opportunity cost of children – other options are getting better (work, play, retire), so having kids is less attractive as in the past.

For more on population projections and the implications to humanity, see an upcoming paper \cite{Dougan2026}

\subsection{United Nations Population Forecasts}

\begin{figure}
    \centering
    \includegraphics[width=1.0\linewidth]{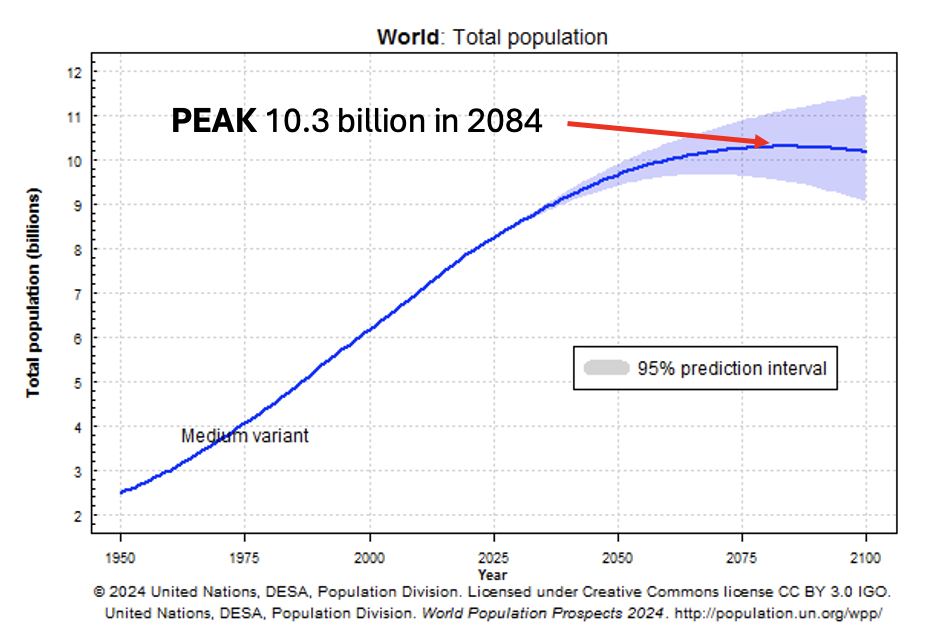}
    \caption{United Nations Forecast 2024}
    \label{fig:UN2024}
\end{figure}

The \cite{UN_WPP_2024} publishes a global population forecast each five years, using a cohort-momentum model projected to the year 2100 (Figure \ref{fig:UN2024}).

According to the UN, population is expected to peak in 2084 at 10.3 billion people.  This is significantly higher than the population in 1950 of 2.5 billion and 1.6 billion in 1900.  This is the power of exponential growth.

\begin{figure}
    \centering
    \includegraphics[width=1\linewidth]{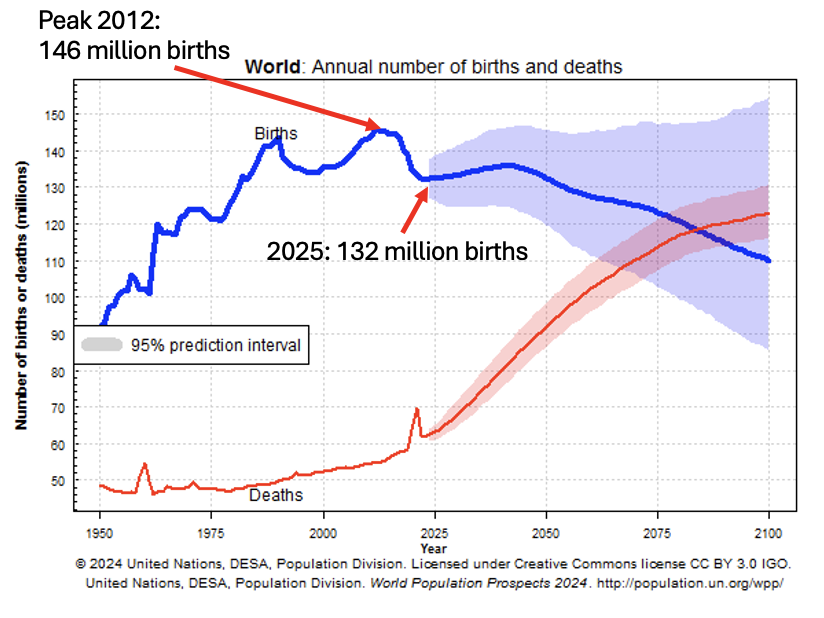}
    \caption{UN Forecast Births and Deaths}
    \label{fig:UNBirths24}
\end{figure}

Taking a closer look at these forecasts shows a large uncertainty around the number of births with less uncertainty about the number of deaths (Figure \ref{fig:UNBirths24}).  Total global births peaked in 2012 at 146 million and has been declining ever since - 132 million births were recorded in 2025.  The UN base case is forecasting an increase in births from 2025 onwards then births decreasing to the current level in 2050.  This increase is speculated to occur due to increased births in sub-saharan Africa, overcoming significant declines in Asia.  This forecast appears misplaced.  In likelihood, births will continue to decline as per the lower bound and global population will peak sometime in the 2060s - in less than 40 years. 

It should be noted that the UN has consistently underestimated population decline.  In 2015, the UN forecast population in 2100 to be 11.2 billion and still growing into the next century.  Five years later in 2019, the UN forecast population to reach 10.9 billion in 2100 and peak sometime in 2100s.  Five years later in 2024, the UN forecast peak population to be 10.3 billion in 2084 and then declining for the first time post 2084. Within 10 years, UN population projections have moved from unlimited growth to peaking and then declining.  So it does not seem unreasonable that the next scheduled UN population forecast in 2029 will most likely show peak population occurring much earlier.  

Furthermore, it is also surprising that the UN and most other NGOs project population only to the year 2100.  This is not long term!  A child born today will live to the year 2100, a year when China’s population will be less than half of what it is today and the world’s population will be declining exponentially.  Why not forecast over longer periods?  This seems unusual.  Some very good long term forecasts have recently been developed by \cite{Spears2024} but more work needs to be done.  

\begin{figure}
    \centering
    \includegraphics[width=1\linewidth]{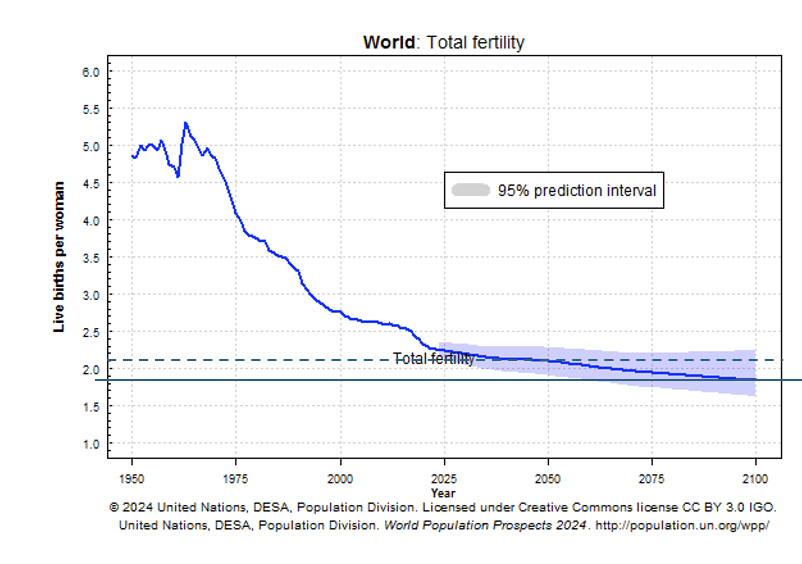}
    \caption{UN TFR Forecast}
    \label{fig:UNTFR24}
\end{figure}

If we review the UN TFR forecasts, we see another unusual forecast - TFR reaching replacement in 2050 (Figure: \ref{fig:UNTFR24}).  Given TFR has rapidly declined over the past 5 years (down from 2.5 to 2.2), it seems unlikely TFR will take another 25 years to decline to 2.1?  Most likely TFR is now below replacement and will continue to decline.  The UN predicts TFR in 2100 at 1.80 and population 10.2 billion.

\begin{figure}
    \centering
    \includegraphics[width=1\linewidth]{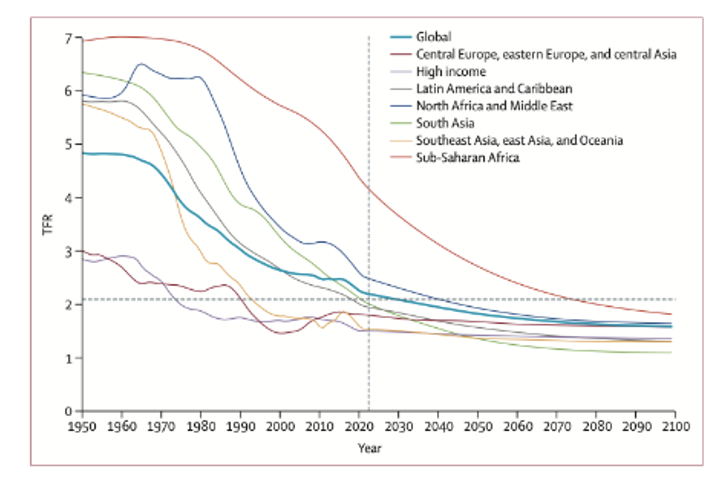}
    \caption{Lancet TFR forecast}
    \label{fig:Lancet24}
\end{figure}

The medical journal The Lancet \cite{Lancet2024} more realistically forecasts TFR to fall to 1.66 in 2100 (Figure \ref{fig:Lancet24}, with population peaking at 9.73 billion in 2064 and falling to 8.79 billion by 2100, some 1.4 billion less than the UN.  To put this difference in context, Lancet predicts global population to be lower than the UN by an amount equal to the entire population of China today.

\subsection{BQI Population Forecasts}

\begin{figure}
    \centering
    \includegraphics[width=1\linewidth]{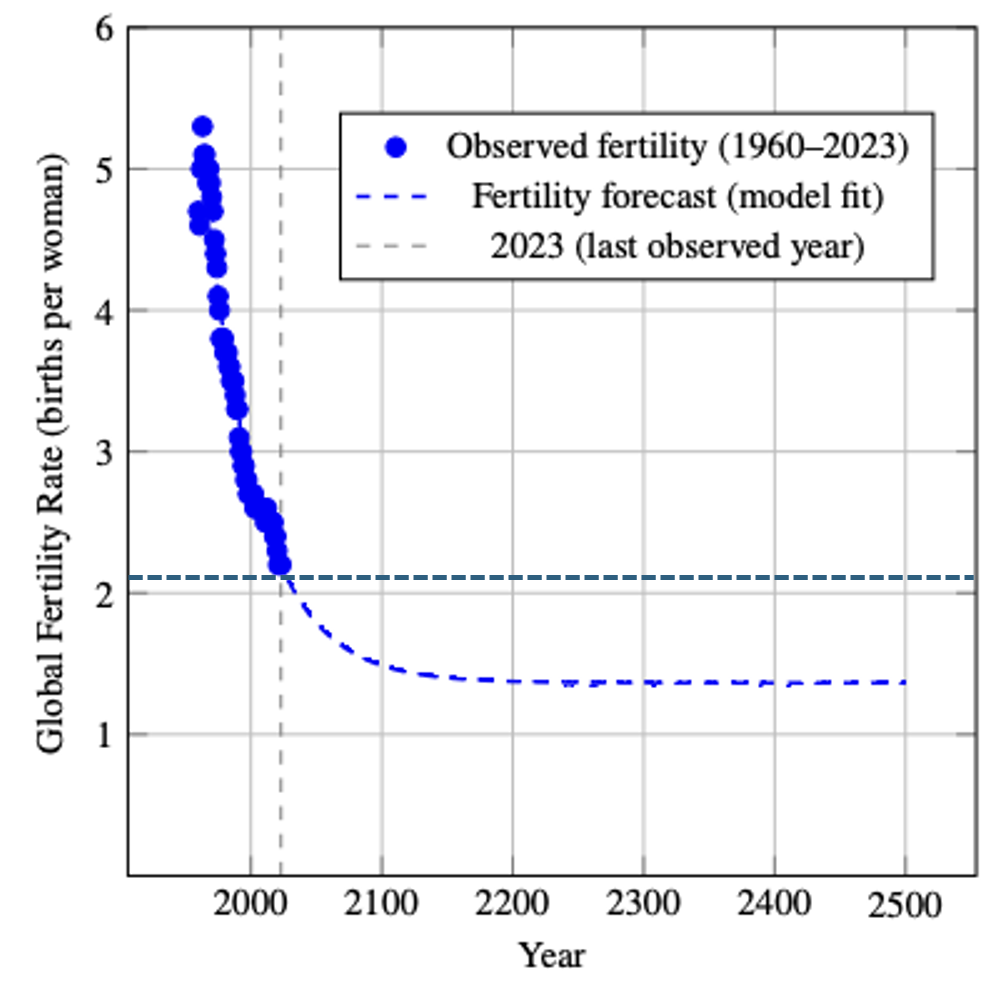}
    \caption{BQI TFR Forecast}
    \label{fig:BQITFR26}
\end{figure}

Fitting a standard regression model to the historic TFR data and extending the forecast timeline to the year 2500, we calculate a fall in the global fertility rate to 1.5 in 2100, then falling further to an asymptotic limit of approximately 1.35 in the 2200s (Figure: \ref{fig:BQITFR26}).  This is a more aggressive decline in TFR compared to the United Nations (1.8 at 2100) and Lancet (1.7 at 2100).   

\begin{figure}
    \centering
    \includegraphics[width=1\linewidth]{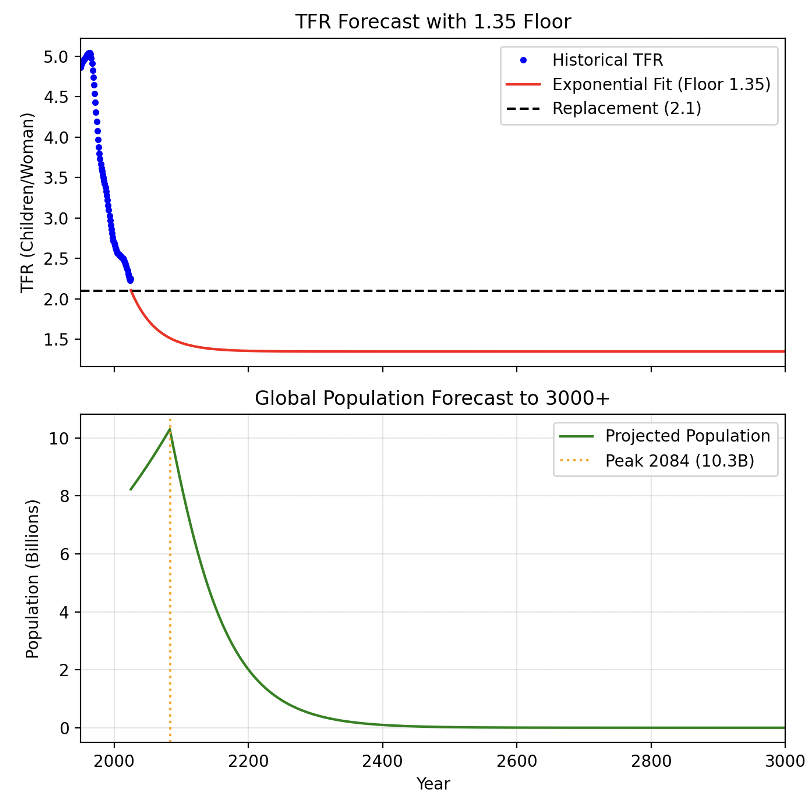}
    \caption{BQI Population Forecast: TFR 1.35 long term}
    \label{fig:BQIPop135}
\end{figure}

Using this TFR profile, we can then calculate population decline using a standard cohort-momentum model.  Figure \ref{fig:BQIPop135} shows the decline.

We see a steep exponential decline after population peaks at 10.3 billion in 2084 (aligned with the UN forecasts), with population declining to 2 billion by 2200 and 1 billion by 2300.  By 2500, global population will be approximately 10 million, with an extinction threshold of less than 10,000 people by the year 3000.  Table \ref{tab:BQIPopTab135} shows the modelling. 

\begin{figure}
    \centering
    \includegraphics[width=1\linewidth]{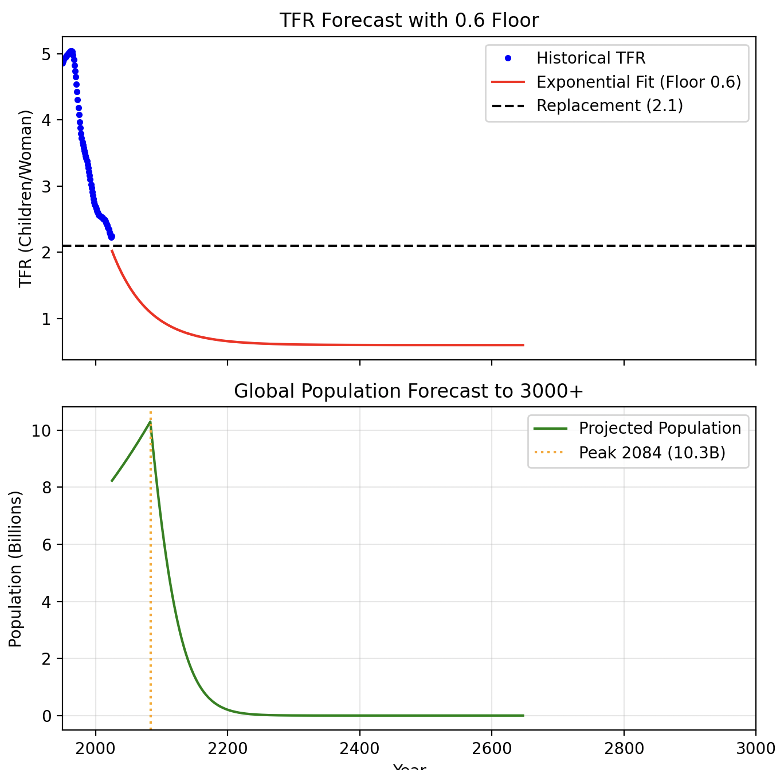}
    \caption{BQI Population Forecast: TFR 0.6 long term}
    \label{fig:BQIPop06}
\end{figure}

In the extreme case where global fertility plateaus at 0.6 (the current TFR of Korea), we see a more rapid rate of decline with extinction within 500 years.  Figure \ref{fig:BQIPop06} and Table \ref{tab:BQIPopTab06} shows an extinction threshold reached in 2434, only 400 years in the future.

It could be argued that TFR falling to 1.35 will not occur, but it is worth noting that no advanced economy has ever reversed a falling TFR trend.  The fact that all of the world except Sub Saharan Africa and Arab countries (both of which also have sharply declining TFR) are now below replacement value of 2.1, shows that 1.35 across the globe is not unrealistic.  Furthermore, there appears no real biological necessity to have above replacement number of births, especially in a society that can consciously decide on the number of offspring.  It should be noted that even if every single woman of child bearing age had either one or two children, extinction would still be inevitable.

\begin{table}
    \centering
    \caption{Forecast population with TFR 1.35 long term}
    \label{tab:BQIPopTab135}
    \begin{tabular}{r r}
        \textbf{Global Population} & \textbf{YEAR} \\
        10,000,000 & 2549 \\
        1,000,000  & 2700 \\
        100,000    & 2851 \\
        10,000     & 3002 \\
        1,000      & 3154 \\
        100        & 3305 \\
        1          & 3607 \\
    \end{tabular}
\end{table}

\begin{table}
    \centering
    \caption{Forecast population with TFR 0.6 long term}
    \label{tab:BQIPopTab06}
    \begin{tabular}{r r}
        \textbf{Global Population} & \textbf{YEAR} \\
        10,000,000 & 2273 \\
        1,000,000  & 2327 \\
        100,000    & 2381 \\
        10,000     & 2434 \\
        1,000      & 2488 \\
        100        & 2541 \\
        1          & 2647 \\
    \end{tabular}
\end{table}

So what is causing the decline in TFR?  There are many causes but none which applies across all cultures and societies. Causes can include:

\begin{itemize} [leftmargin= 0.5cm]
    \item Access to contraception
    \item Economic and affordability concerns
    \item Lower rates of marriage
    \item Digitization \& less socialization
    \item Future environmental and geopolitical concerns
    \item Higher levels of women’s education \& career participation
    
\end{itemize}

It is clear that women are having children later in life so biologically this implies women will have fewer children.  Furthermore, the opportunity cost of having children is increasing as other options for couples continues to improve.  Better work and career options, better leisure and travel opportunities, and better lifestyle options makes having children less attractive.

\subsection{Implications for Humanity}

Depopulation could be categorized under the economic term "Tragedy of the Commons".  The tragedy of the commons occurs when individuals act in their own self-interest which depletes a shared resource which is not in the best long term interests of the species.  Climate change is one example of the Tragedy of the Commons.  Depopulation is likely a much more significant tragedy, as individuals act in their own self interest and have fewer offspring, thereby depleting a limited resource (offspring) which inevitably leads to extinction of the species.

Depopulation is clearly a significant issue for humanity and one that is seldom discussed in the general media.  The United Nations and other government agencies only forecast out to 2100, not further.  However, as TFR continues to plummet across all economies, humanity will need to face an existential crisis of ageing and depopulation in the following centuries.  Fewer people has significant implications for society and the future of humanity, implications which have yet to be comprehensively debated and analysed.  A brief list of these implications is given below.
\\

\noindent \textbf{Positive implications:}
\begin{itemize} [align=right, leftmargin=*]
    \item fewer people reduces global emissions and diminishes the threat of climate change 
    \item fewer people reduces resource consumption
    \item fewer people minimises encroachment into the environment with less pollution and a rewilding of the earth
    \item fewer people should lessen territorial disputes leading to less wars and conflict
    \item fewer people will lower the chances of pandemics given less people to transmit (although higher potential of extinction if low population)
\end{itemize}

\noindent \textbf{Negative implications:}
\begin{itemize} [align=right, leftmargin=*]
    \item fewer people to create literature, music, science, physics, art etc will negatively impact innovation, productivity and novelty
    \item fewer people will lead to negative GDP growth resulting in lower standards of living
    \item fewer younger people will be required to provide taxation to support older generations
    \item fewer people to provide healthcare and assistance to elderly populations
\end{itemize}

But what of economic incentives to have more babies?  China is a case point where this policy has been applied to no avail.  Once a society reaches a level of wealth and prosperity that reduces TFR, it can't be reversed with economic coercion.  Economic coercion simply pulls forward decisions that would likely already have been made.

\begin{figure}
    \centering
    \includegraphics[width=1\linewidth]{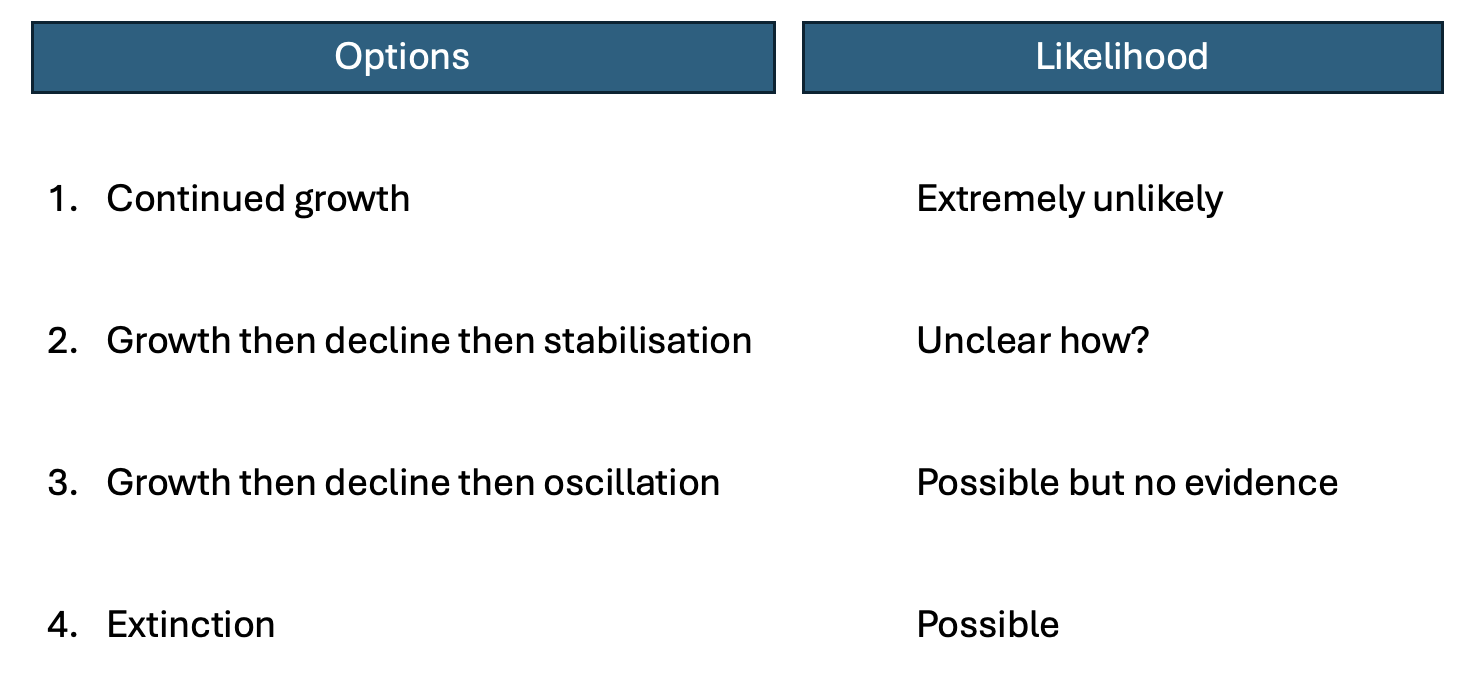}
    \caption{Future Population Options}
    \label{fig:HumanityOptions}
\end{figure}

\begin{figure}
    \centering
    \includegraphics[width=1\linewidth]{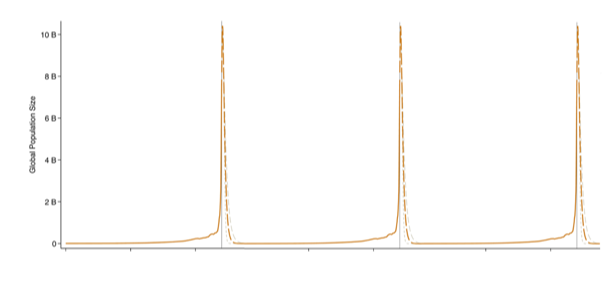}
    \caption{Oscillating Population (adapted from \protect\cite{Spears2024})}
    \label{fig:Oscillating}
\end{figure}

So what are the options facing humanity?  Figure \ref{fig:HumanityOptions} lists possible scenarios.  The first is continued growth past 10.3 billion in 2084.  This seems extremely improbable given current fertility rates and should be discounted.  The second option is that fertility rates plateau at replacement level sometime in the future.  Again, this seems unlikely unless some major event forces society to address population decline.  It's unclear what this would be.  Another possible outcome is an oscillating population as shown in Figure \ref{fig:Oscillating}.  In this scenario, as population declines the world becomes worse and the opportunity cost of having children decreases, thereby increasing TFR and subsequently population growth restarts.  This is possible but there is no evidence to support this claim.  And this option assumes the world's population suffers reduced standards of living, leading to repopulation.  But machines and AI will inevitably lead to stable outcomes for humanity, thus reinforcing the Tragedy of the Commons (see Section \ref{sec:Machines} which discusses machines and AI in more detail).

As humanity's extinction draws nearer, there will be inevitable discussions about preserving the human race.  Cloning, gene manipulation, living longer, cryogenics, and other scientifically motivated \textit{"Brave New World"} solutions will be explored.  However, ethically, the human race will rebel against artificial population growth and depopulation will continue its exponential decline.  Hence, the only plausible scenario of the four listed above is extinction.\\

\section{Exodemography}

\subsection{Exodepopulation}
If depopulation leading to an extinction event is a possibility for humanity, could this be a universal phenomenon for all intelligent life forms?  Is species depopulation a possible solution to the Fermi paradox and a new Great Filter?  Exospecies depopulation might occur given four key factors:

\begin{enumerate} [label=\arabic*., align=right, leftmargin=*]
    \item Exospecies is at the top of the food chain, reducing predation and the need for large numbers of offspring
    \item Exospecies develops sufficient technology to improve infant mortality
    \item Exospecies learns how to control reproduction via contraception and abortion
    \item Exospecies reaches level of wellbeing whereby procreation becomes a lower priority
\end{enumerate}

If we assume that Darwinian evolution is prevalent across the universe and that all life must evolve in a Darwinian way, then Factor 1 seems to be self evident.  An extra terrestrial life form would need to evolve to the top of the food chain in order to subsequently develop a civilisation.  So at that point, the number of offspring would not be determined by predation and would tend to a lower natural number.

On earth, apex predators such as primates, lions, sharks, dolphins, whales and seals all have less offspring, reproduce more slowly and invest more heavily in parental care.  The reasons for this include lack of predation, energy scarcity, “quality over quantity”, and population regulation of prey.  For an exospecies, lower numbers of offspring would apply independent of the form of sexual reproduction, whether it be asexual, sexual or something else.  Any species having evolved via Darwinian evolution would have lower numbers of offspring compared to other species on its planet.

If we look at Factor 2, technology clearly reduces infant mortality.  Exospecies with intelligence would presumably develop health care – medicines, vaccinations, neonatal care etc. which would reduce infant mortality, leading to fewer births required.  For humanity, this has been a large driver of reduced fertility rates as the number of births per woman needed to bear children who survive to adulthood has been falling.  Note Benjamin Franklin's observation back in 1755 which assumed 50\% infant mortality rate \footnote{Benjamin Franklin himself had three biological children to two women, of which two survived to adulthood (33\% mortality rate)}.  An intelligent exospecies would also use technology to improve its environmental situation, enhancing its food and water sources, improving its sanitation and waste management, and providing better nutrition to its young.  All factors that would improve infant mortality.

Factor 3 is a significant advance on reducing offspring.  Humanity is the only species on planet earth that has developed technology to consciously stop reproduction.  It should be noted that this is not the same as some species who have sex for fun (eg dolphins, bonobos) or species who delay reproducing due to resource constraints (eg kangaroos, rats, meerkats).  Humans have invented technology to ensure sex at any time does not lead to pregnancy.  An exospecies with sufficient technology would be highly likely to develop contraception, thereby further reducing fertility rates.  Moreover, an exospecies would likely develop safe abortion procedures to further reduce the number of offspring born.

The final Factor 4 is the most controversial.  Just like humanity, would an exocivilization develop a society which is wealthy enough that the cost of offspring is outweighed by other benefits?  We would assume successful exocivilizations would have a high standard of living given its technology and environment, and that as alien life improves, bearing offspring would compete with other alien life options.  So alien birth rates would then fall further, just like humanity

Defining an Exospecies Fertility Rate ("EFR"), we come to the conclusion that likely the EFR would decline just like our TFR, due to the four Factors listed above.

\subsection{Depopulation - a new Great Filter}

\begin{figure}
    \centering
    \includegraphics[width=1\linewidth]{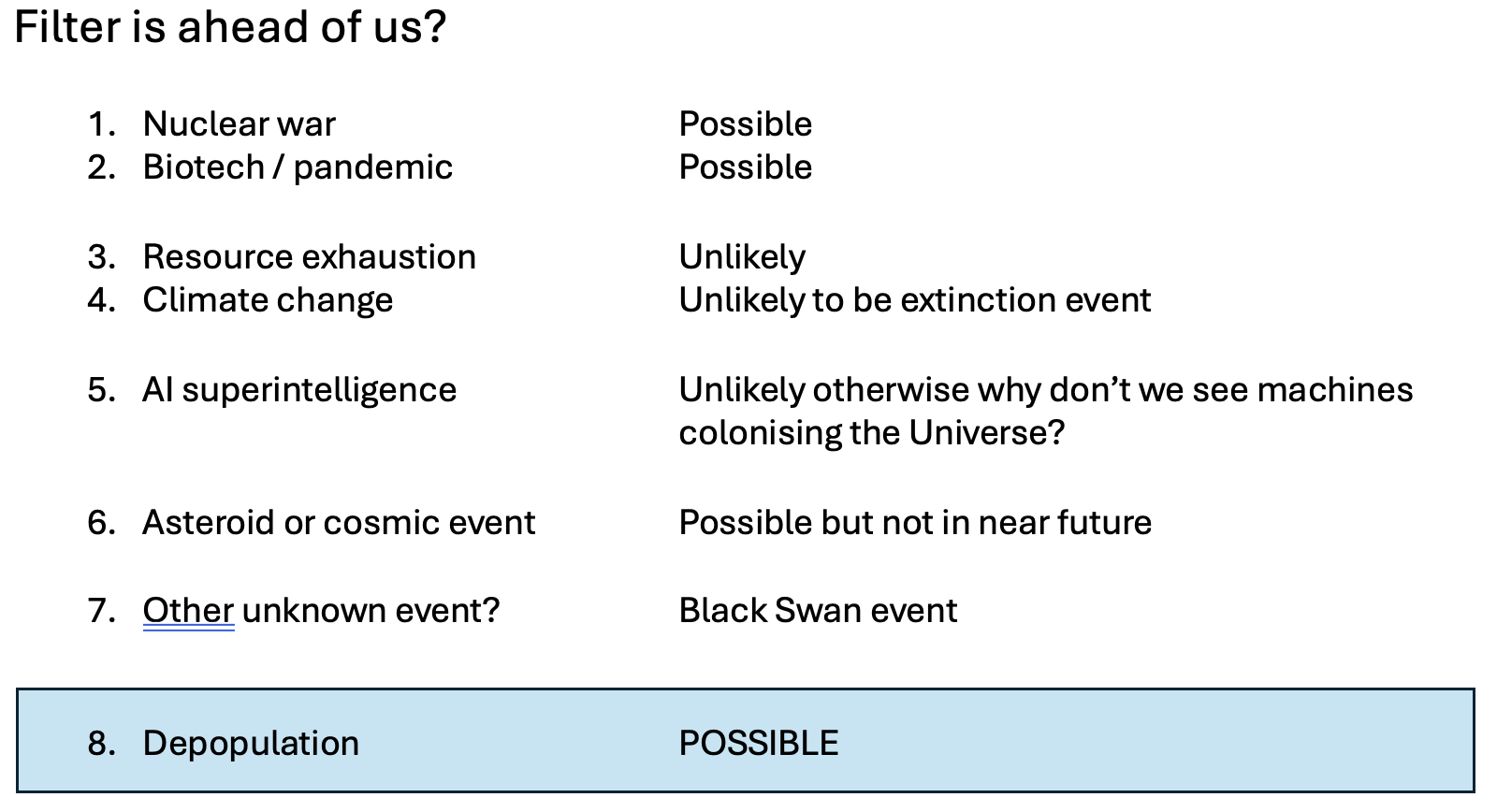}
    \caption{Possible Filters in front of humanity}
    \label{fig:GreatFilters}
\end{figure}

In summary, unlike a cosmic event or asteroid impact, depopulation could lead to a "quiet, benign extinction event" where a species simply ceases to replace itself due to shifting social, economic, or psychological values.

Of course, this is a very anthropogenic argument, but no more than say nuclear weapon annihilation or biological weapon extinction, or even AI extinction.  See Figure \ref{fig:GreatFilters} for a lists of possible future extinction events.

In any case, we can't rule out the possibility that depopulation "for whatever reason" in fact causes intelligent species to extinct themselves, this rationale being no different from the assumption that advanced exocivilizations will destroy themselves with nuclear technology, another very anthropogenic assumption.

Or, if not an extinction, a rapidly ageing or shrinking population may lose the "critical mass" of innovators and labourers required to maintain complex technology or launch energy-intensive interstellar colonization efforts.

As exocivilizations become more technologically advanced and wealthy, birth rates could fall below replacement levels. If this trend is universal for all intelligent life, it could be a fundamental barrier to galactic expansion.

\section{Corollary - Machine consciousness} \label{sec:Machines}

A natural response to the depopulation problem is that as the population of the world declines and ages, machines and AIs will replace humans in many areas of service and production.  This is already occurring in places like Korea and Japan where technology is replacing labour, and of course the transformational growth of AI will undoubtedly supplement human effort.

So surely an exospecies will also develop machines, AI and other technologies to continue expansion?  But if this is so, the Fermi Paradox can be restated as \textit{"Why don't we see machines and AIs colonizing the galaxy?"}.

The answer may be that machines \textbf{will not become conscious} and will continue to be tools for humanity to use.  The Fermi Paradox can then be restated as \textbf{"Machines will not become conscious, otherwise we would see them colonising the universe"}.

If machines were to become conscious, then this would further add to the depopulation problem, as humans decline in numbers and machines take over the planet. However, this seems unlikely given the absence of evidence of this evolution occurring elsewhere in the universe.  More likely, machines will continue to do the work that humans can't do, or perform work humans don't want to do.  When the last human perishes, just like the last Tasmanian Tiger which died in captivity in a zoo with no mates to procreate with; when this occurs for humanity sometime in the not too distant future, machines will simply no longer be needed and will cease to operate.  Then our civilization will become extinct, just like other exocivilizations who follow this same path.

\section{Conclusion}
This paper highlights three key takeaways.

\begin{enumerate} [label=\arabic*., align=right, leftmargin=*]
    \item Depopulation is a potential extinction event for humanity
    \item Depopulation dynamics (exodemography) could apply to exocivilizations
    \item Depopulation may be a Great Filter lying in front of both humanity and all intelligent life forms
\end{enumerate}

There is no doubt that depopulation as a Great Filter deserves to be addressed.  This paper is the first to discuss at a high level the implications of depopulation not only to humanity, but also to exocivilizations.  It is not unreasonable to assume that humans will become the first species on this planet to consciously extinct ourselves, depopulating ourselves out of existence in a benign, slow way.  As Hemingway once remarked on bankruptcy, "It happens slowly, then all at once", and so too might humanity slowly move towards extinction and then all of sudden, in an exponential way, become extinct.

If benign human extinction is even a remote possibility, then it must be considered as a possible resolution of the Fermi Paradox and subsequently a Great Filter we have yet to pass through.

\section*{Acknowledgements}

The author wishes to thank Prof. Mikhail Prokopenko and Prof. Dean Rickles, both at the University of Sydney, and Prof. Paul Davies at Arizona State University, for their helpful comments and constructive feedback on an earlier version of this work.  The author also acknowledges the participants of the University of Sydney Heron Island Complexity Conference 2026 for valuable and lively discussions on the depopulation topic.

%%%%%%%%%%%%%%%%%%%%%%%%%%%%%%%%%%%%%%%%%%%%%%%%%%



\bibliographystyle{mnras}
\bibliography{GreatFilter.bib} % if your bibtex file is called DD.bib

@article{Lancet2024,
  author  = {{GBD 2021 Fertility and Forecasting Collaborators}},
  title   = {Global fertility in 204 countries and territories, 1950--2021, with forecasts to 2100: a comprehensive demographic analysis for the Global Burden of Disease Study 2021},
  journal = {The Lancet},
  year    = {2024},
  volume  = {403},
  number  = {10440},
  pages   = {2057--2099},
  month   = {May},
  doi     = {10.1016/S0140-6736(24)00550-6},
  url     = {https://doi.org},
  note    = {Published online March 20, 2024}
}

@techreport{UN_WPP_2024,
  author      = {{United Nations, Department of Economic and Social Affairs, Population Division}},
  title       = {World Population Prospects 2024: Summary of Results},
  institution = {United Nations},
  year        = {2024},
  type        = {UN DESA/POP/2024/TR/NO. 9},
  address     = {New York},
  url         = {https://population.un.org}
}

@unpublished{Dougan2026,
  author = {Darren J. Dougan},
  title  = {Depopulation},
  year   = {2026},
  note   = {Preprint available on arXiv:2601.12345 [cs.LG]},
  url    = {https://arxiv.org}
}

@article{Carter1983Anthropic,
  author    = {Carter, Brandon},
  title     = {The Anthropic Principle and its Implications for Biological Evolution},
  journal   = {Philosophical Transactions of the Royal Society of London. Series A, Mathematical and Physical Sciences},
  volume    = {310},
  number    = {1512},
  pages     = {347--363},
  year      = {1983},
  publisher = {The Royal Society},
  url       = {https://doi.org}
}

@incollection{Franklin1755Observations,
  author    = {Franklin, Benjamin},
  title     = {Observations Concerning the Increase of Mankind, Peopling of Countries, etc.},
  year      = {1755},
  booktitle = {Observations on the Late and Present Conduct of the French, with Regard to Their Encroachments upon the British Colonies in North America},
  editor    = {Clarke, William},
  publisher = {S. Kneeland},
  address   = {Boston},
  note      = {Written in 1751},
  pages     = {Appendix}
}

@article{Spears2024,
    doi = {10.1371/journal.pone.0298190},
    author = {Spears, Dean AND Vyas, Sangita AND Weston, Gage AND Geruso, Michael},
    journal = {PLOS ONE},
    publisher = {Public Library of Science},
    title = {Long-term population projections: Scenarios of low or rebounding fertility},
    year = {2024},
    month = {04},
    volume = {19},
    url = {https://doi.org/10.1371/journal.pone.0298190},
    pages = {1-16},
    number = {4},

}

@misc{WorldFertility2023,
  author       = {WorldBank},
  title        = {WorldFertility},
  howpublished = {On-line data},
  url          = {https://data.worldbank.org/indicator/SP.DYN.TFRT.IN?end=2023&start=1960&view=chart},
  year         = {2023},
  note         = {Accessed: 2025-10-12}
}

@article{Safuan2012_predator_prey,
  author       = {Hamizah M. Safuan and H. S. Sidhu and Z. Jovanoski and I. N. Towers},
  title        = {A two‑species predator‑prey model in an environment enriched by a biotic resource},
  journal      = {ANZIAM Journal},
  volume       = {54},
  year         = {2012},
  pages        = {C768--C787},
  doi          = {10.21914/anziamj.v54i0.6376},
  note         = {Received 9 November 2012; revised 20 June 2014}
}

@article{BasenerRoss2005,
  author       = {Bill Basener and David S. Ross},
  title        = {Booming and Crashing Populations and Easter Island},
  journal      = {SIAM Journal on Applied Mathematics},
  volume       = {65},
  number       = {2},
  pages        = {684--701},
  year         = {2005},
  doi          = {10.1137/S0036139903426952},
  note         = {Mathematical model (system of ODEs) capturing resource‑harvest vs resource regeneration leading to overshoot and collapse}
}

@article{Frank2018,
  author       = {Adam Frank and Jonathan Carroll--Nellenback and Martina Alberti and Axel Kleidon},
  title        = {The Anthropocene Generalized: Evolution of Exo‑Civilizations and Their Planetary Feedback},
  journal      = {Astrobiology},
  year         = {2018},
  volume       = {18},
  number       = {5},
  pages        = {561--578},
  doi          = {10.1089/ast.2017.1671},
  note         = {Includes a generalized civilization–planet feedback model, drawing analogy to systems such as Easter Island}  
}

@incollection{Drake1975,
  author    = {Frank D. Drake and Carl Sagan},
  title     = {The Search for Extraterrestrial Intelligence},
  booktitle = {Current Aspects of Exobiology},
  editor    = {G. Mamikunian and M. H. Briggs},
  publisher = {Pergamon Press},
  year      = {1975},
  pages     = {323--345},
  address   = {New York},
  note      = {Originally based on the formulation presented by Drake in 1961},
}

@misc{Hanson1998,
  author       = {Robin Hanson},
  title        = {The Great Filter — Are We Almost Past It?},
  howpublished = {On-line essay},
  note         = {Originally published 1996, updated Sept.\ 15, 1998},
  url          = {http://mason.gmu.edu/~rhanson/greatfilter.html},
  year         = {1998}
}

% \begin{thebibliography}{99}
% \bibitem[\protect\citeauthoryear{Author}{2012}]{Author2012}
% Author A.~N., 2013, Journal of Improbable Astronomy, 1, 1
% \bibitem[\protect\citeauthoryear{Others}{2013}]{Others2013}
% Others S., 2012, Journal of Interesting Stuff, 17, 198
% \end{thebibliography}

%%%%%%%%%%%%%%%%%%%%%%%%%%%%%%%%%%%%%%%%%%%%%%%%%%

%%%%%%%%%%%%%%%%% APPENDICES %%%%%%%%%%%%%%%%%%%%%

% \appendix

% \section{TBD}

%%%%%%%%%%%%%%%%%%%%%%%%%%%%%%%%%%%%%%%%%%%%%%%%%%

% Don't change these lines
\bsp	% typesetting comment
\label{lastpage}
\end{document}